\documentclass[9pt,twocolumn,twoside]{opticajnl}
\journal{opticajournal} 

\setboolean{shortarticle}{false}

\usepackage{lineno}

\newcommand{\red}[1]{{\color{black}#1}}

\title{Narrow-linewidth exciton-polariton laser}

\author[1]{Bianca Rae Fabricante}
\author[1]{Mateusz~Kr\'ol}
\author[1]{Matthias Wurdack}
\author[2]{Maciej Pieczarka}
\author[3]{Mark Steger}
\author[3]{David W. Snoke}
\author[4]{Kenneth West}
\author[4]{Loren N. Pfeiffer}
\author[5]{Andrew G. Truscott}
\author[1]{Elena A. Ostrovskaya}
\author[1,*]{Eliezer Estrecho}

\affil[1]{ARC Centre of Excellence in Future Low-Energy Electronics Technologies and Department of Quantum Science and Technology, Research School of Physics, The Australian National University, Canberra ACT 2601 Australia}
\affil[2]{Department of Experimental Physics, Faculty of Fundamental Problems of Technology, Wrocław University of Science and Technology, Wyb. Wyspianskiego 27, 50-370 Wrocław, Poland }
\affil[3]{Department of Physics and Astronomy, University of Pittsburgh, Pittsburgh, Pennsylvania 15260, USA}
\affil[4]{Department of Electrical Engineering, Princeton University, Princeton, New Jersey 08544, USA }
\affil[5]{Department of Quantum Science and Technology, Research School of Physics, The Australian National University, Canberra ACT 2601 Australia}

\affil[*]{eliezer.estrecho@anu.edu.au}

\begin{abstract}
\red{Exciton-polariton laser is a promising source of coherent light for low-energy applications due to its low-threshold operation. However, a detailed experimental study of its spectral purity, which directly affects its coherence properties is still missing. Here}, we present a high-resolution spectroscopic investigation of the energy and linewidth of an exciton-polariton laser in the single-mode regime, which derives its coherent emission from an optically pumped and confined exciton-polariton condensate. We report an ultra-narrow linewidth of 56~MHz or 0.24~$\mu$eV, corresponding to a coherence time of 5.7~ns. The narrow linewidth is consistently achieved by using an exciton-polariton condensate with a high photonic content confined in an optically induced trap. Contrary to previous studies, we show that the excitonic reservoir created by the pump and responsible for creating the trap does not strongly affect the emission linewidth as long as the condensate is trapped and the pump power is well above the condensation (lasing) threshold. \red{The long coherence time of the exciton-polariton system uncovered here opens up opportunities for manipulating its macroscopic quantum state, which is essential for applications in classical and quantum computing.}
\end{abstract}

\setboolean{displaycopyright}{false} 

\begin{document}

\maketitle

\section{Introduction}

The spectral linewidth or the inverse coherence time of a laser is an important measure of its spectral purity. For photonic lasers, the theoretical limit is set by the Schawlow--Townes (ST) linewidth~\cite{schawlow_infrared_1958}, which arises from phase fluctuations caused by quantum noise, and is inversely proportional to the number of particles (photons) $N_0$ in the lasing mode. For matter wave lasers, e.g. coherent atom beams, where particle interactions play a significant role, the theoretical limit is broadened due to particle collisions that turn number fluctuations into energy or frequency fluctuations~\cite{thomsen_atom-laser_2002}. In general, interparticle interactions lead to linewidth broadening that scales with $N_0$ in a class of lasers with a Kerr-like  nonlinearity, such as atom and polariton lasers.

Polariton lasers are coherent light sources generated by the decay of bosonic condensates of exciton-polaritons (polaritons for short) ~\cite{kasprzak2006bose, Balili1007}, hybrid particles arising from the strong coupling of photons and excitons (electron-hole pairs). These lasers, typically built using an optical microcavity with embedded exciton-hosting material, achieve lasing operation without population inversion~\cite{imamog1996nonequilibrium} and at a threshold much lower than that of an equivalent conventional photon laser~\cite{deng2003polariton}. Their \red{low-threshold} operation, hybrid light-matter nature, large nonlinearity, and solid-state platform are promising for a number of low-power applications, such as ultra-fast optical polarization switches, modulation applications, compact sources of terahertz radiation, and logic elements~\cite{Zhang2022,kavokin_baumberg_malpuech_laussy,Sanvitto2016}. Polariton condensates also find applications in both classical and quantum computing~\cite{kavokin2022polariton} where a long coherence time and a narrow linewidth of the polariton emission can play an important role.

\begin{figure}[ht!]
\centering
\includegraphics[width=\linewidth]{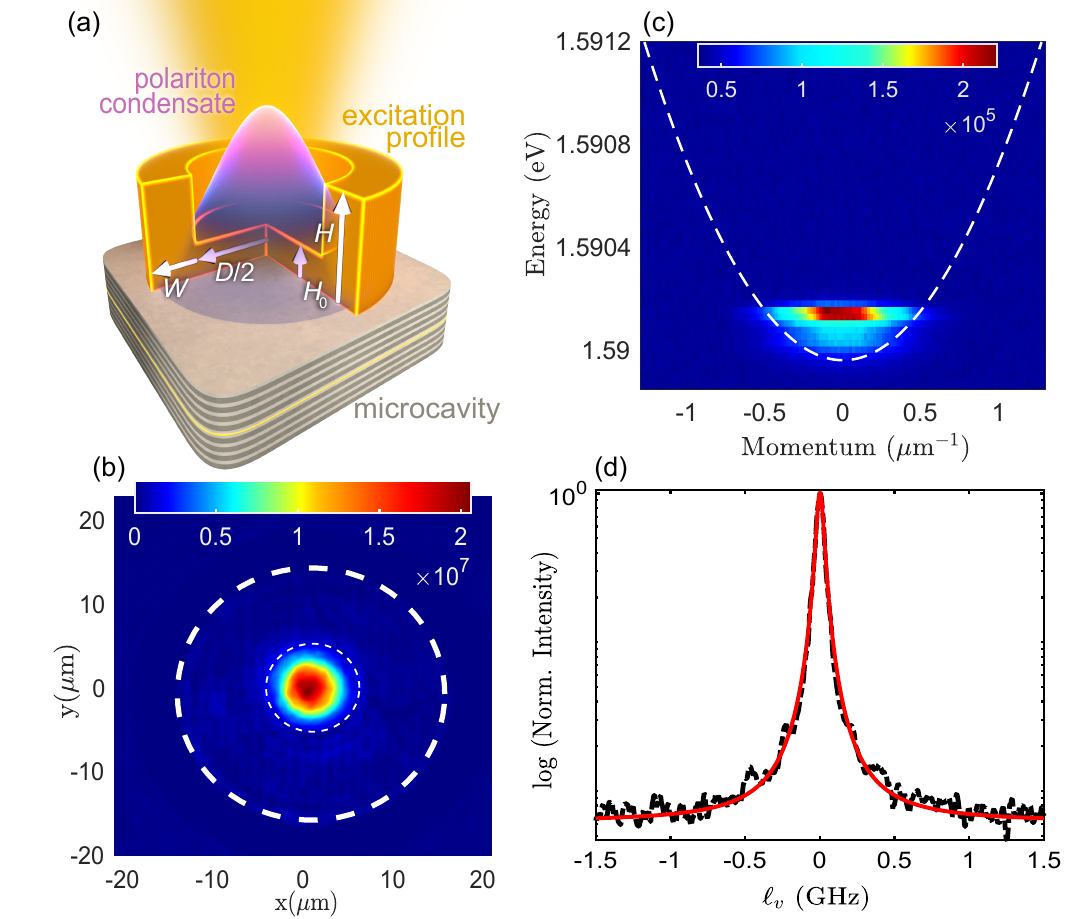}
\caption{Experimental configuration. (a) Schematics of the experiment showing the excitation laser profile with parameters height $H$, inner diameter $D$ and inner offset $H_0$, which results in a similarly-shaped optically-induced trap $\mathcal{V}$ for the polariton condensate. (b) Example real space condensate emission (bright emission) inside the trap created by the pump laser (dashed line corresponds to the edges of the pump). (c) Momentum-space emission of the single-mode condensate. \red{The dashed line is the polariton dispersion extracted from low-power measurement (see Supplemental Document).} (d) Example spectrum (black dots) \red{of the emission of the condensate or the polariton laser} measured using the scanning Fabry-P\'erot Interferometer (SFPI) and fitted to a Lorentzian function \red{(red solid line)} with a FWHM of $\ell_\nu=56$~MHz. \red{There is no other peak within the free spectral range of the SFPI (see Supplemental Document). Here, the excitation and condensate emission powers are $48.6$~mW and 60 $\mu$W, respectively.}}
\label{fig:1}
\end{figure}

Theoretical investigations of the polariton laser linewidth~\cite{tassone2000lasing, porras2003linewidth, amelio2022bogoliubov} predicted the interplay between the ST mechanism and the self-phase modulation arising from particle interactions. The ST mechanism narrows the spectrum as the number of condensate particles $N_0$ increases at low densities. At high densities, the interaction energy $UN_0$, where $U$ is the interaction strength, dominates and leads to linewidth broadening with increasing $N_0$. Hence, an optimal $N_0$ with the narrowest linewidth~\cite{porras2003linewidth} is predicted to occur close to the condensation threshold. However, a motional narrowing is predicted at high densities when fluctuations decay much faster than the coherence time~\cite{Whittaker_2009}, negating the broadening induced by interactions.

In reality, polariton lasers are more complicated since polaritons coexist with an incoherent excitonic reservoir that repels and feeds the condensate~\cite{byrnes2014exciton}, resulting in an additional decoherence mechanism arising from number fluctuations in the reservoir. This can potentially impose a lower limit on the linewidth. In the case of optical pumping, both reservoir and condensate particle numbers are directly affected by the intensity of the pump laser, making the pump the main source of fluctuations. A narrow linewidth corresponding to a coherence time of 150~ps was revealed in Ref.~\cite{Love2008}, using a single-mode excitation laser instead of multimode ones used previously~\cite{kasprzak2006bose, Balili1007}. This achievement was later followed by a demonstration of a shot-noise-limited intensity-stabilized polariton laser using a mode-selective microcavity~\cite{Kim2016} with a relatively short 60-ps coherence time.
In both cases, the excitonic reservoir, created by the excitation laser, directly overlapped with the condensate.


An order-of-magnitude enhancement of the coherence time was achieved by minimising the condensate--reservoir overlap using optically induced traps. In Ref.~\cite{Askitopoulos2019}, the interference visibility persisted beyond 1~ns with an extrapolated coherence time of up to 3~ns when the condensate is trapped, compared to $<$100~ps when the condensate is ``untrapped'' {but localized by gain~\cite{Wouters2008, ostrovskaya2012dissipative, roumpos2010gain}, with a tail expanding away from the excitation spot}. The prolonged coherence time of trapped condensates enabled detection of fine structure splitting of the polariton condensate energy~\cite{orfanakis_ultralong_2021}, self-induced Larmor precession~\cite{sigurdsson_persistent_2022}, and long-lived coherently coupled counter-circulating currents~\cite{barrat_qubit_2023}. However, a systematic study of the role of the reservoir on the decoherence of polaritons is incomplete.


Previous experiments on polariton coherence typically rely on a Michelson interferometer to measure the first-order correlation function $g^{(1)}$. This technique images the interferogram at different time delays to construct $g^{(1)}(\tau)$, where $\tau$ is the time delay. This requires multiple realizations of the experiment and averaging over timescales orders of magnitude longer than the polariton lifetime and the coherence time, which unfortunately, washes out fluctuations in energy and linewidth. Furthermore, a precise motorized delay stage is required, but its physical size becomes impractical for coherence times longer than 1 ns.

In this work, we use a scanning Fabry-P\'erot interferometer (SFPI) to directly measure the linewidth and small energy shifts of chopped continuous-wave (quasi-CW pulses) polariton condensates~\cite{kuznetsov2023microcavity}. The unprecedented resolution enabled by the SFPI reveals the fluctuations in energy and linewidth within a single quasi-CW condensate pulse. We present the successful measurement of a narrow Lorentzian linewidth of 56~MHz, which is {four times} narrower than the previous record~\cite{Askitopoulos2019}, translating to a long coherence time of 5.7~ns. We further investigate how the narrow linewidth can be achieved by probing the influence of the optical trapping on the linewidth (coherence time) and reveal that the decoherence effects arising from the overlap between the condensate and the excitonic reservoir are only significant in the near-threshold regime. For larger excitation powers above the threshold, our polariton laser is highly robust against decoherence as long as the condensate is trapped.

\section{Excitation profile and linewidth measurement}

We use a ring-shaped excitation profile to create and trap the condensate~\cite{askitopoulos2013polariton, estrecho2019direct, sun2017direct}, which has been an established trapping configuration for studying the fundamental properties of polariton condensates, such as the temporal coherence~\cite{Askitopoulos2019}, thermalization~\cite{SunPolLifetime, alnatah2024coherence}, excitation spectra~\cite{pieczarka2022bogoliubov, bieganska2021collective, pieczarka2020observation, estrecho2021low}, and critical fluctuations~\cite{alnatah2024critical}. As schematically drawn in Fig.~\ref{fig:1}(a), the excitation profile has an annular shape, with inner diameter $D$, wall thickness $W$, wall height $H$, and inner offset $H_0$.
The resulting optically induced potential $\mathcal{V}$ assumes a similar shape but with `soft' edges due to the excitonic reservoir diffusion ~\cite{askitopoulos2013polariton}.

We use a spatial light modulator (SLM) from Meadowlark Optics in amplitude configuration to turn a broad Gaussian laser beam into the desired shape, then image it onto the sample using a high-NA objective. The excitation laser is a single-mode Ti:Sapphire laser (Coherent 899-21), chopped by an acousto-optic modulator (AOM) to 8-ms quasi-CW pulses every 800 ms, and then coupled to a single-mode polarization-maintaining fiber. The excitation laser wavelength is around 719~nm, tuned to the lowest reflectance minimum of the microcavity, while the polariton lasing wavelength is around 780~nm. 



\begin{figure*}[hbt!]
    \center
    \includegraphics[width=\textwidth]{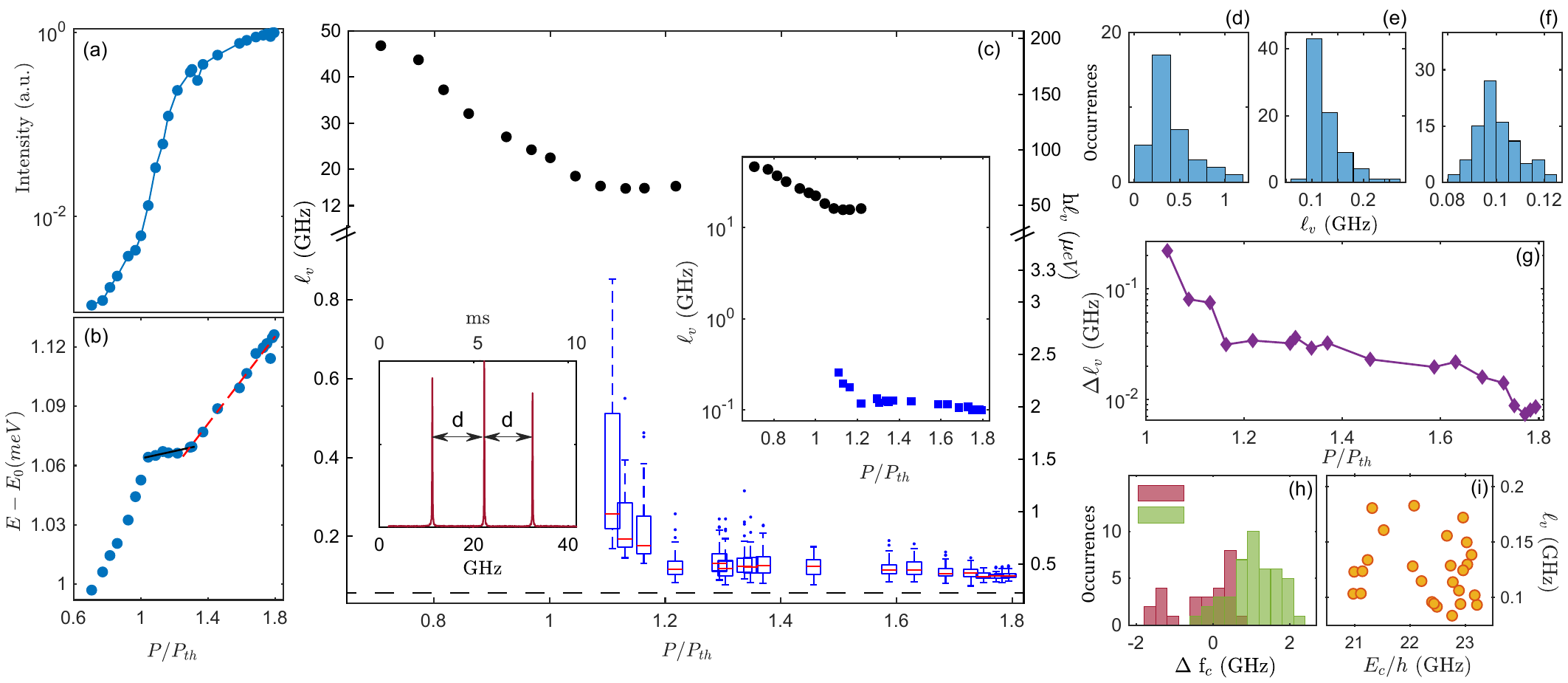}
    \caption{Pump-power dependence. (a) Intensity and (b) energy of the condensate as a function of normalized excitation power, $P/P_{th}$. The slopes of the linear fits in (b) are 0.0185~meV(or 4.35~GHz)$/(P/P_{th})$ and 0.111~meV(or 26.8~GHz)$/(P/P_{th})$ for the black and red lines, respectively. (c) Linewidth of the polariton emission as a function of $P/P_{th}$ measured using (black dots) the spectrometer and (blue box plots) the SFPI. The box plots show the interquartile range (length of box), the median (red line), and the whole range (dashed lines). The dots are considered outliers (see {SI} for details). Left inset: Example measured signal from the SFPI of a single quasi-CW condensate pulse scanned over 3 FSRs \red{of the SFPI}. Right inset: Combined measurements in log scale. (d-f) Distribution of the measured linewidths at $P=$1.04 $P_{th}$, 1.16 $P_{th}$, and 1.80 $ P_{th}$, respectively. The standard deviations are (d) 150~MHz, (e) 40~MHz, and (f) 8~MHz. (g) Standard deviation of linewidth, $\Delta\ell_{v}$. (h) Energy distributions at $P=1.16 P/P_{th}$ measured (red) between shots, centered with respect to the median, i.e. $E-E_{med}$, and (green) within the shot between two FSRs, i.e. $E-E_{FSR}$. The standard deviations are (red) $0.8~GHz$ and (green) $0.64$~GHz. (i) Scattered plot of simultaneous linewidth and energy measurements corresponding to (f). \red{For this dataset, $P_{th}\approx 30$~mW.}
    }
    \label{fig:2}
\end{figure*}

The sample is a high Q-factor GaAs/AlGaAs planar microcavity with a cavity photon lifetime$\tau_{C}\sim$150~ps~\cite{Steger_polaritonlifetime}, enclosed in a cryostat at a temperature of $<$10~K. We restrict the measurements to a region of the sample where polaritons have low excitonic (or high photonic) content at an excitonic fraction of $|X|^2$=$ 0.13$, corresponding to a detuning of $\delta $=$ -17$ meV and a polariton lifetime of $\sim175~ps$ (see {SI} for more details).
We use the low excitonic fraction to minimise strong linewidth broadening due to interactions \cite{porras2003linewidth, thomsen_atom-laser_2002, Kim2016, amelio2022bogoliubov} while allowing for the formation of a single-mode confined condensate in the ground state. If the excitonic fraction is too low, the condensate tends to occupy several high-energy states of the trap~\cite{estrecho2019direct,KRIZHANOVSKII2001583,Tassone1997}. An example of polariton condensate photoluminescence (PL) emission is presented in Fig.~\ref{fig:1}(b) showing the confined condensate in the center of the ring-shape potential barrier created by the excitation laser. The corresponding energy and momentum distribution is presented in Fig.~\ref{fig:1}(c) showing a single-mode ground-state emission.

We use a SFPI from Thorlabs (SA210-5B) with a 10-GHz free spectral range (FSR) to measure the spectrum of the single-mode polariton laser (see \red{Supplemental Document} for details).  We synchronized the chopping of our laser and the scanning of the Fabry-P\'erot such that at least 3 FSRs are spanned in every single quasi-CW pulse of the condensate.
An example of the measured spectrum fitted to a Lorentzian function is presented in Fig.~\ref{fig:1}(d).

The normalized power spectrum $I(\omega)$, where $\omega/2\pi = \nu$ is the frequency, is related to the first-order coherence function $g^{(1)}$ by the Fourier transform~\cite{loudon2000quantum}:
\begin{equation}
    I(\omega) = \frac{1}{2\pi} \int_{-\infty}^\infty g^{(1)}(\tau) e^{-\textrm{i}\omega\tau} d\tau,
    \label{eq:I}
\end{equation}
with a corresponding coherence time, $\tau_{coh}$, defined as~\cite{loudon2000quantum}:
\begin{equation}
    \tau_{coh} = \int_{-\infty}^\infty |g^{(1)}(\tau)|^2 d\tau.
    \label{eq:tauc}
\end{equation}
If $g^{(1)}$ is exponential of the form $\exp(-\gamma\tau)$, the coherence time is $\tau_{coh}=1/\gamma$, and $I(\nu)$ is Lorentzian, as in our case, with a full width at half maximum (FWHM) or spectral linewidth of $\ell_\nu=1/(\pi\tau_{coh})$.
If $g^{(1)}$ is a Gaussian of the form $\exp(-\tau^2/2s^2)$, the coherence time is $\tau_{coh}=\sqrt{\pi}s$, and the linewidth is $\ell_\nu=\sqrt{2\ln{2}\pi}/(\pi\tau_{coh})$. We use these formulas to compare linewidths and coherence times in the Discussion section.

\section{Excitation power dependence}
We first study the dependence of the spectral properties of the polariton emission on the excitation power, $P$, relative to the lasing threshold $P_{th}$ defined by the onset of nonlinear increase in photoluminescence intensity with increasing excitation power, as shown in Fig.~\ref{fig:2}(a).

\begin{figure*}[h!]
    \centering
    \includegraphics[width=\textwidth]{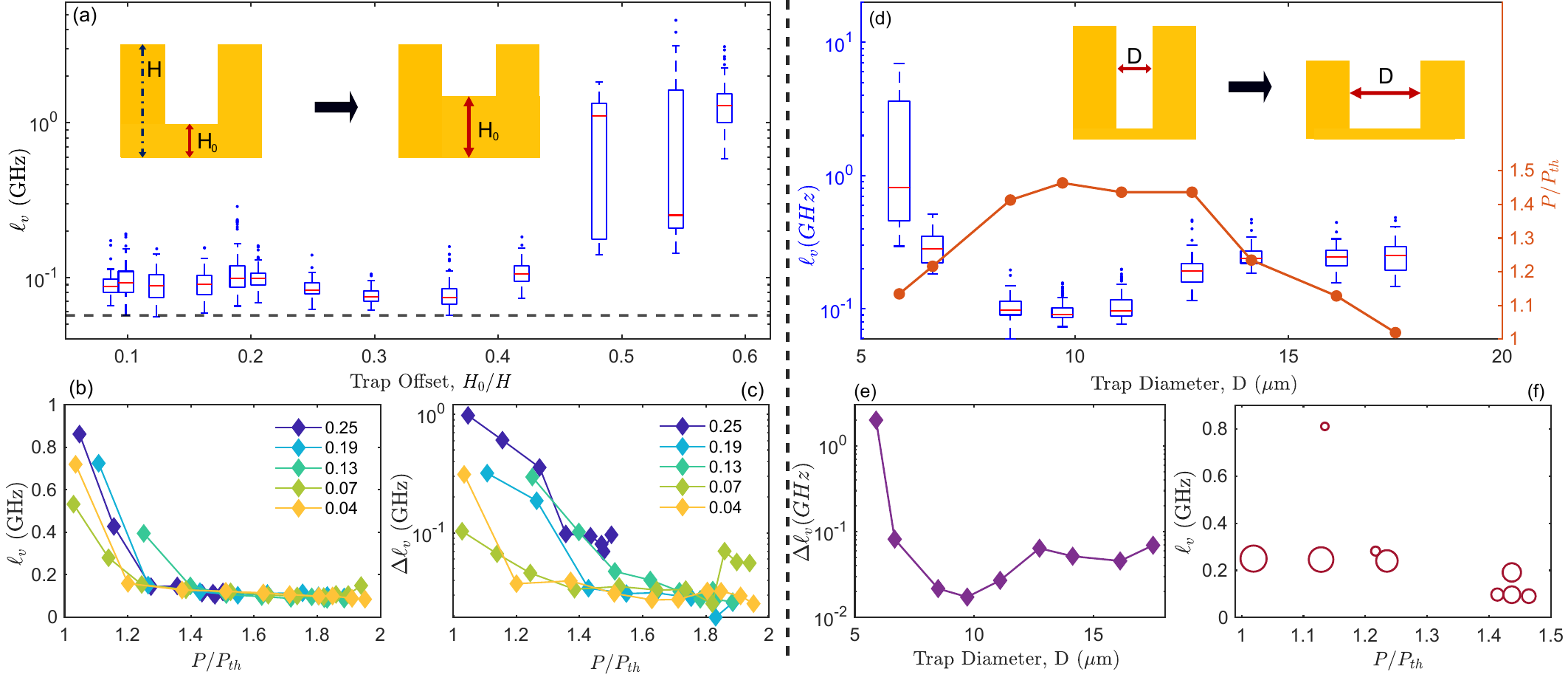}
        \caption{Probing the overlap with the reservoir. (a) Linewidths as a function of excitation inner offset $H_{0}$, with $H$ fixed and $D = 9~\mu$m. (b) Median and (c)  standard deviation $\Delta$$\ell_{v}$ of measured linewidths as a function of normalized excitation power $(P/P_{th})$ for various $H_{0}/H$. (d) Linewidths and (e) their standard deviation $\Delta\ell_\nu$ as a function of trap diameter $D$. The right y-axis in (d) denotes the corresponding normalized excitation power $P/P_{th}$ as the trap diameter is tuned. (f) Median linewidth plotted as a function of normalized excitation power. Marker size represents the trap diameter. The insets in (a) and (d) show how the excitation profile changes as $H_0$ and $D$ are tuned, respectively.}
    \label{fig:3}
\end{figure*}

The condensate energy $E$ grows with $P$, as shown in Fig.~\ref{fig:2}(b), which is expected for polariton condensates. Different contributions to this blueshift can be described using the following approximation:
\begin{equation}\label{eq:EP}
    E(P) = E_0 + \mathcal{T}_0(P) + g_R n_{R,0}(P) + gn_{c,0}(P),
\end{equation}
where $E_0$ is the polariton energy at zero momentum in the low-density limit. The second term $\mathcal{T}_0(P)$ is the quantum confinement energy of the optically induced trap, which increases with $P$ since the trap gets tighter {and potential walls higher} with increasing $P$~\cite{Pieczarka2019}. The last two terms originate from interparticle interactions, where $g$ ($g_R$) is the polariton--polariton (polariton--reservoir) interaction strength and $n_{c,0}$ ($n_{R,0}$) is the condensate (reservoir) density around the center of the trap. Both the reservoir and condensate densities grows with $P$, such that at high excitation powers, $P>P_{th}$, they dominate over $\mathcal{T}_0$. Note that the trapping potential is created by the reservoir distribution, i.e. $\mathcal{V}(x,y)=g_Rn_R(x,y)$. However, we simplify it here such that $\mathcal{V}(x,y)=\mathcal{T}_0+\mathcal{V}_0$, where $\mathcal{T}_0$ captures the shape of the potential barrier and $\mathcal{V}_{0}=g_Rn_{R,0}$ captures the energy shifts imposed by the trap offset $\mathcal{V}_0$ due to non-negligible reservoir density at the bottom of the trap. Note that $\mathcal{V}_{0}$ is proportional to the inner offset $H_0$ of the excitation profile, while $\mathcal{T}_0$ is strongly affected by the trap diameter $D$. Equation~(\ref{eq:EP}) clearly suggests that any fluctuation in both the reservoir $n_R$ and condensate $n_c$ densities leads to energy fluctuation of the condensate, and hence to the linewidth broadening.


The dependence of the linewidth on the excitation power is presented in Fig.~\ref{fig:2}(c), which shows a remarkable three-order-of-magnitude linewidth narrowing across the threshold $P_{th}$. This linewidth narrowing is one of the hallmark signatures of polariton condensation \cite{kasprzak2006bose,deng2010exciton}, which signals the emergence of long-range order, or in this case long temporal coherence. The  black dots are measured using a conventional spectrometer showing a spectral narrowing from 50 GHz (200~$\mu$eV) down to its resolution limit of 0.025~nm or 12~GHz (50~$\mu$eV).

To probe narrower linewidths, we use a SFPI with a resolution of 56 MHz. With this spectral resolution, we can clearly see the fluctuations in energy and linewidth both between and within shots, as shown by the distributions in Fig.~\ref{fig:2}(d-i,). A shot is defined as a single pulsed condensate created by the quasi-CW excitation pulse and an example measurement is shown in the left inset of Fig.~\ref{fig:2}(c).
Care is taken in analyzing the spectral profiles due to the long time scale of the measurement ($>10^{-3}$~s) relative to the coherence time ($10^{-9}$~s). In most cases, the measured spectrum is symmetric and can be fitted to a Lorentzian, such as in Fig.~\ref{fig:1}(d), representing situations where the optical spectrum did not substantially change during the measurement. 

The resulting distributions of measured linewidths are presented in Fig.~\ref{fig:2}(c) using a box plot to simultaneously show the asymmetry and spread of the data. The plot clearly shows that the linewidth continues to narrow as the excitation power is increased. {For $P/P_{th}>1.2$, the linewidths plateau to a median of $\approx$100~MHz up to the maximum excitation power of $P_{max}/P_{th} \approx 1.8$}. This trend is accompanied by a narrowing of the spread of the shot-to-shot variations of the linewidth, down to a standard deviation of $\Delta\ell_\nu=8~MHz$, as shown in Fig.~\ref{fig:2}(g). The latter suggests that the linewidth is becoming less sensitive to fluctuations at higher excitation powers. As shown in the next section, the same trend persists as we change other parameters of the trap that have a direct effect on the overlap of the condensate and the reservoir. {Note that close to the maximum power of $P/P_{th}=1.8$, the distribution of linewidths are relatively symmetric [see Fig.~\ref{fig:2}(f)] and the fluctuation in linewidth is not correlated to the fluctuation in energy [see Fig.~\ref{fig:2}(i)].}

We highlight that many shots have resolution-limited linewidths, shown by the lower whiskers of the box plots reaching the resolution limit (dashed horizontal line) in Fig.~\ref{fig:2}(c). This is also supported by the asymmetric distribution of the linewidths, which are truncated by the resolution limit, e.g. see Fig.~\ref{fig:2}(e).  This suggests that the actual linewidth in some shots could go well below 50~MHz or a coherence time $\tau_{coh}$ towards 10s of ns. This possibility will be further explored in future experiments using a higher resolution SFPI or other interferometric techniques, such as delayed self-heterodyne detection~\cite{bai2021narrow}.



\section{Dependence on trap parameters}
We change the parameters of the excitation profile [see Fig.~\ref{fig:1}(a)], and hence the $n_R(x,y)$ distribution, to probe how the different terms in Eq.~(\ref{eq:EP}) affect the linewidth. This effectively changes the relative strengths of the different terms in the equation.

We first tune the inner offset $H_0$ while keeping all other parameters of the excitation profile fixed (see \red{Supplemental Documen} for more details). This offset is proportional to the third term ($\mathcal{V}_0=g_{R,0}n_R$) in Eq.~(\ref{eq:EP}) and will not strongly affect the other terms. Hence, this measurement directly probes the influence of the reservoir on the polariton laser linewidth. The dependence of the linewidth on $H_0$ at the maximum excitation laser power is presented in Fig.~\ref{fig:3}(a). Interestingly, the offset does not significantly affect the linewidth ($\sim$~100 MHz) until a certain value, at which the linewidth drastically broadens to $\sim$~1 GHz. Note that increasing $H_0$ also increases the excitation power $P$. However, the observed dependence on $H_0$ is the reverse of the $P/P_{th}$ dependence observed in Fig.~\ref{fig:2}(c), where the linewidth drastically decreases with $P$. This is because changing $H_0$ unavoidably changes $P_{th}$.

To compare the results on an equal footing, we scanned $P$ for different offsets $H_0$ to determine $P_{th}$ and the results are presented in Fig.~\ref{fig:3}(b) using the normalized excitation power $P/P_{th}$. The trend clearly shows a general behavior consistent across different offsets: the linewidth narrows with increasing $P/P_{th}$ and plateaus to $\sim$100~MHz at $P/P_{th}>1.2$. It now matches the power dependence in Fig.~\ref{fig:2}(c). However, larger offsets tend to have larger linewidths at the same normalized excitation power, especially for $P/P_{th}<1.2$. In addition, the linewidth variations, $\Delta\ell_\nu$, shown in Fig.~\ref{fig:3}(c), decrease and plateau in a similar manner but at $P/P_{th}>1.5$. These findings strongly suggest that the overlap between the excitonic reservoir and condensate only plays a significant role in linewidth broadening near the condensation threshold, i.e. for $P<1.2P_{th}$. At higher excitation powers, the trend is the same regardless of the trap offset (at least within our spectral resolution), and the linewidth is robust against power fluctuations. {This could be due to the hole-burning effect~\cite{estrecho2018single, estrecho2019direct} that results in lower than expected trap offsets \textcolor{red}{${H}_0$} at higher excitation powers, regardless of the excitation offset $H_0$.
}

zzz{Note that tuning $H_0/H$ towards large values also allows us to probe the interplay between trapping and localization by gain~\cite{amelio2022bogoliubov}, since the trapping becomes less effective for large offsets ($H_0/H>0.45$) such that an ``untrapped'' condensate~\cite{ostrovskaya2012dissipative} can form at $H_0/H \rightarrow 1$. In previous experiments~\cite{Askitopoulos2019}, untrapped condensates tends to have order-of-magnitude larger linewidths (short coherence times) than trapped ones. Hence, the significantly large linewidths observed for large offsets ($H_0/H>0.45$) can be also be attributed to the increased role of gain~\cite{amelio2022bogoliubov}.}

We also tune the inner diameter $D$ of the trap while fixing the excitation intensity, the wall width $W$, and inner offset ($H_0=0$). Note, however, that the parameter \textcolor{red}{$H$}, which is proportional to the potential height, changes with $D$ since the amplitude mask will sample different radial distances from the center of the input Gaussian beam [see Fig.~\ref{fig:3}(d) inset]. Changing $D$ will mainly affect $\mathcal{T}_0$ in Eq.~(\ref{eq:EP}), since the confinement energy is inversely proportional to the trap area. The change will also affect $n_{R,0}$ but this effect will only be strong when $D$ is small, i.e. when the center of the trap is closer to the walls. In both cases, we expect the energy fluctuation to be stronger, and hence linewidths broader, for smaller $D$.

The measured linewidths are presented in Fig.~\ref{fig:3}(d) for different trap diameters. The plot clearly show a non-monotonic behavior. At small $D$ ($<8~\mu$m), the median and the spread of the measured linewidths are quite large ($\sim$1~GHz), as expected. Increasing $D$ drastically narrows down the median linewidth to $<$100 MHz and then slowly broadens it again to around 250 MHz. The spread in measured linewidths behave in a similar manner, as shown in Fig.~\ref{fig:3}(e).

The non-monotonic behavior arises from the unavoidable change of the excitation power $P$ and condensation threshold $P_{th}$ as $D$ is varied. The normalized excitation power $P/P_{th}$, plotted as orange dots in Fig.~\ref{fig:3}(d), clearly shows the same (albeit inverted) non-monotonic behavior. This provides further evidence that the linewidth tends to broaden close to threshold. However, for the same $P/P_{th}$, smaller traps tend to have larger median linewidths [see Fig.~\ref{fig:3}(f)]. For example, for the similar ratio of $P/P_{th}\approx1.1$, the condensate in the 6-$\mu$m trap has a median $\ell_\nu\approx1$~Ghz,  which is an order of magnitude larger than that in the 16-${\mu m}$ trap. This behavior is  similar to Fig.~\ref{fig:3}(b), further suggesting that the overlap with the reservoir, which is large for smaller traps, strongly broadens the linewidth at excitation powers close to threshold.

{In addition to changing $P_{th}$, increasing the trap size can promote multi-mode behavior due to increased occupation of the excited states as the level spacing decreases. Spontaneous emission into these states effectively results in additional noise to the lasing mode (the ground state)~\cite{amelio2022bogoliubov}, which can contribute to the observed broadening for $D>12\mu$m.}

\section{Fluctuations and time scale analysis}

\red{Finally, we analyze the fluctuation in the condensate energy $E$ and its relation to the measured linewidth $\ell_\nu$. Fluctuation in energy leads to linewidth broadening, but the energy fluctuation and linewidth measurements occur at different timescales. It takes about 2.5 ms to scan over a single FSR or to measure any energy shift, but it only takes $<50~\mu$s to scan over a single narrow spectral peak. Hence, we can probe effects of slow and fast noise by probing the fluctuation in $E$ and the linewidth, respectively.}

\red{We first look at the slow fluctuation by analyzing the distribution of the energy shifts, such as the one shown in Fig.~\ref{fig:2}(h), exhibiting a mean shift of $\approx1.0$~GHz. The mean shift can arise from a slow increase in excitation power within the 8-ms quasi-CW pulse due to a slow increase in AOM efficiency after turning on. Heating of the sample due to the excitation laser can also contribute, but would result in a redshift instead of the blueshift observed here.}

\red{The fluctuation above the mean energy shift, quantified by the standard deviation $\Delta E$, is $\Delta E=0.64$~GHz within the shot (every 2.5 ms) and $\Delta E=0.80$~GHz from shot to shot (every 800 ms), for the dataset in Fig.~\ref{fig:2}(h). The former is generally smaller than the latter such that for each power in Fig.~\ref{fig:2}(c) the average is around $\Delta E\approx1.0$~GHz and $\Delta E\approx1.6$~GHz, respectively.  This means that there is less energy drift (or less fluctuation) on the ms time scale compared to longer time scale. Furthermore, the energy fluctuation does not have a strong dependence on the excitation power, trap offset, and ring diameter (see Supplementary Document for the relevant plots).}

\red{Assuming the excitation power is the only source of noise,} we can estimate the \red{equivalent fluctuations in $P$} responsible for the observed energy fluctuation using the slope of the linear fits on the experimental $E(P)$ in Fig.~\ref{fig:2}(b). Fluctuation \red{of} $\Delta E\approx1$~ GHz correspond to power fluctuations of 22\% and 4\% for $P/P_{th}<1.3$ and $P/P_{th}>1.3$ in Fig.~\ref{fig:2}(b), respectively. The latter is comparable to the measured intensity fluctuations of the laser in this timescale. The former is quite large and can arise from the other \red{sources of technical} noise such as SLM flicker, air currents, and mechanical vibrations of the cryostat, which can all lead to large intensity modulation and \red{shifts the excitation spot on the sample. The latter can change the excitation profile and the polariton detuning resulting in changing $P_{th}$ and polariton energy  $E_0$ (see Eq.~\ref{eq:EP}), cumulatively changing the condensate energy $E$ even if the input power $P$ is the same.} An active feedback can be implemented using the AOM to modulate the excitation intensity to minimize the energy drift at these timescales.





\begin{figure}[t!]
    \centering
    \includegraphics[width=\columnwidth]{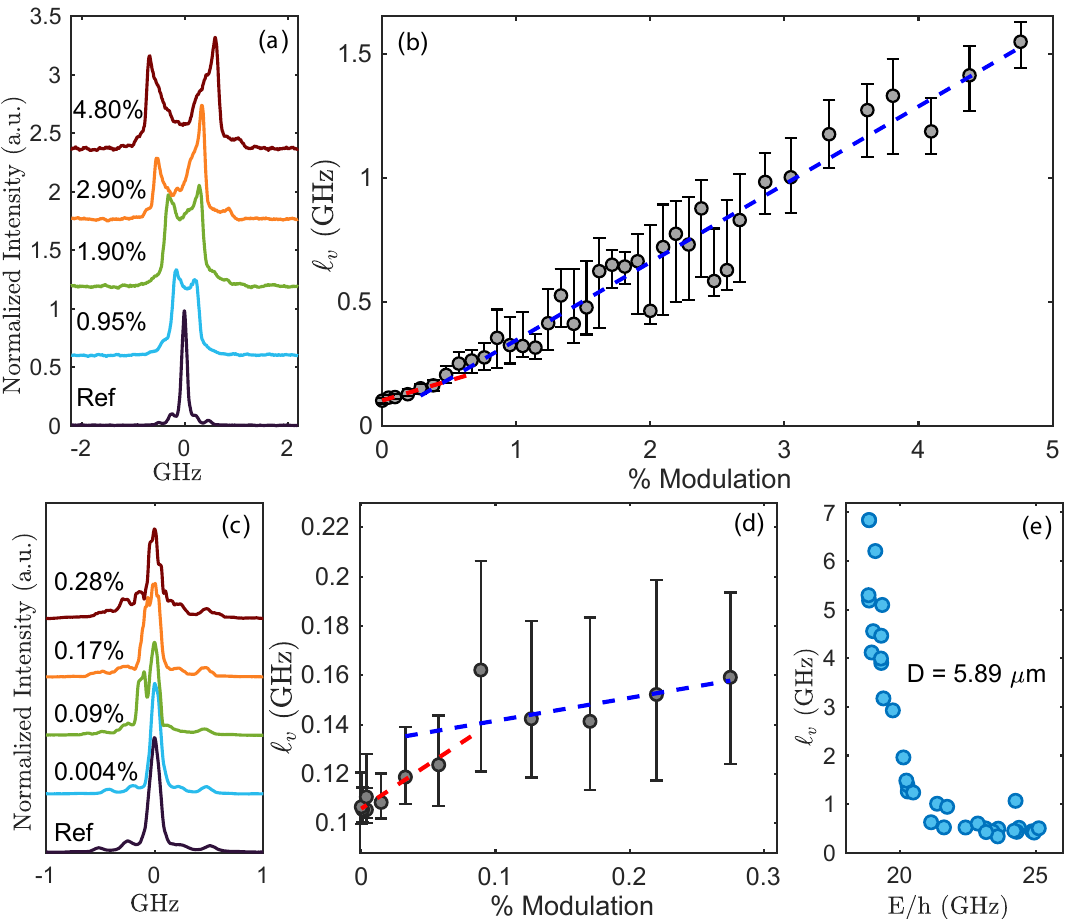}
    \caption{Effect of external modulation on the linewidth measurement.
    (a) Spectral profiles and (b) effective linewidths measured for increasing strengths of 1-MHz sinusoidal modulation. (c) Spectral profile and (d) effective linewidths measured for increasing strength of added Gaussian noise. (e) E-$\ell_{v}$ correlation for smallest trap size in Fig.~\ref{fig:3}.}
    \label{fig:4}
\end{figure}



To probe the faster timescales ($<10^{-4}$~s), we analyze the linewidth by intentionally \red{modulating} the excitation laser intensity using the AOM to externally introduce controllable fluctuations. Sample spectra are presented in Fig.~\ref{fig:4}(a) for different amplitudes of a 1-MHz sinusoidal modulation. The results for low frequencies are presented in the {SI}. The lineshape of the spectra are clearly affected by the added modulation such that it deviates from a Lorentzian to a flat-top and eventually towards a two-peak profile. We define the effective linewidth using the FWHM of the spectra and their dependence on the modulation strength is presented in Fig.~\ref{fig:4}(b), clearly showing the broadening of the spectra with increasing modulation strength. This suggests that if fast modulations in $P$ or in the reservoir density are present, they naturally lead to broadening of the linewidth.

We also introduce Gaussian noise to the laser intensity to emulate realistic random noise across all frequency range (at least those allowed by the AOM). The sample spectra and the modulation amplitude dependence are presented in Fig.~\ref{fig:4}(c,d), respectively. Instead of flat-top-like spectra, random fine peaks appear in the spectra as the modulation strength is increased. Note however, that the ``envelope" over these peaks resembles a more Gaussian-like distribution, e.g. around 0.14\%. Below 0.05\% peak-to-peak random noise, the measured linewidth remains the same as in the non-modulated case. This puts an upper limit (of approximately the same level) of noise that may exist in the experiment.

Finally, we demonstrate the extreme sensitivity of the linewidth on small shot-to-shot fluctuations in energy $E$ or power $P$ close to the threshold. The simultaneous energy and linewidth measurements taken from single quasi-CW pulses of the condensate are plotted in Fig.~\ref{fig:4}(e) for the experimental conditions corresponding to the 6-$\mu$m trap in Fig.~\ref{fig:3}(d), which has an average $P/P_{th}\approx1.1$. The plot clearly shows that if the energy blueshifts by a small amount $\sim$1~GHz, the linewidth narrows by an order of magnitude.  This means that the large measured fluctuations in linewidth $\Delta\ell_\nu$ close to the threshold are due to this extreme sensitivity. At higher $P/P_{th}$, when the linewidth plateaus, the $E$-$\ell_\nu$ correlation vanishes [see for example Fig.~\ref{fig:2}(i)] and the measured $\Delta\ell_\nu$ reflects the uncorrelated fluctuations in linewidth. 



\section{Discussions}

We presented precise linewidth and energy shift measurements of the laser emission from an optically trapped single-mode polariton condensate using a scanning Fabry-P\'erot interferometer. We found three general behaviors regarding the linewidth. First, the linewidth drastically narrows with increasing excitation power and plateaus to a median of $\ell_\nu<100$~MHz \red{(measured within 50~$\mu$s)} for $P/P_{th}>1.2$. This behavior is independent of the overlap with the reservoir, as confirmed by increasing either the trap offset or the trap size, as long as the condensate is trapped. Second, the reservoir only strongly affects the linewidth at excitation powers close to threshold ($P/P_{th}<1.2$), where larger overlap leads to larger linewidths. Third, the fluctuations in the measured linewidth decrease with excitation power down to $\Delta\ell_{v}=8$~Mhz at $P/P_{th}\approx1.8$. Hence, to optimize the linewidth, the normalized excitation power should be increased beyond $P/P_{th}>1.2$, followed by reduction of the overlap of the condensate with the reservoir.

To compare our results with previous works, we converted the reported coherence times to linewidths using Eqs.~(\ref{eq:I}) and (\ref{eq:tauc}). The narrowest linewidth we measured ($\ell_\nu=56$~MHz) is four times narrower than the previous record~\cite{Askitopoulos2019}, but note that the latter was only extrapolated from a limited dataset. In terms of coherence time, the Lorentzian lineshape in our case translates to $\tau_{coh}=1/{\pi\ell_\nu} = 5.7$~ns, which is twice longer than in Ref.~\cite{Askitopoulos2019}. The comparison with other reported values~\cite{kuznetsov2023microcavity,Love2008,Kim2016,barrat_qubit_2023,orfanakis_ultralong_2021, Askitopoulos2019} of coherence time and linewidths are tabulated in the \red{Supplemental Document}.

We can also compare the linewidth of the polariton laser to that of vertical cavity surface emitting photonic laser (VCSELs), which share the same sample structure but is not affected by particle interactions. Our linewidth is on par with the state-of-the-art single-mode VCSELs (50~MHz)~\cite{serkland2020narrow,huang2022narrow}, despite the presence of strong nonlinearity in polariton lasers. Better design, such as in multi-mirror VCSELs~\cite{huang2022narrow}, as well as electrical injection of carriers~\cite{bhattacharya2014room, schneider2013electrically}, can further narrow the linewidth.



We expect the true linewidth of our polariton laser to be less than 50~MHz, beyond the resolution limit in our experiment, as evidenced by the skewed distribution in Fig.~\ref{fig:2}(f) and the low-amplitude external modulation measurements in Fig.~\ref{fig:4}. This will signify the coherence time up to the order of 10-100 of ns, paving the way for coherent manipulation of condensates at experimentally accessible time scales, for example, in realizing quantum computations with polariton qubits~\cite{xue2021split}. Increasing the linewidth resolution using either a higher-resolution SFPI or other interferometric techniques~\cite{bai2021narrow} will be essential in finding the optimum excitation conditions~\cite{porras2003linewidth} for obtaining the narrowest linewidth and the longest coherence times.

Note that the case of higher excitation powers ($P/P_{th}>2$) still needs to be examined. Theory predicts the linewidth to broaden again at high excitation power due to self-interaction and fluctuations in particle number ~\cite{porras2003linewidth,Haug}. This is supported by multiple experiments where linewidths broaden at high excitation powers above threshold~\cite{kasprzak2006bose, askitopoulos2013polariton, Kim2016, deng2003polariton, Balili1007}. However, a motional narrowing of the linewidth, which can occur when the fluctuations are much faster than the coherence time, is predicted to occur at high excitation powers ~\cite{Whittaker_2009}. This may occur at an intermediate pump power regime, which we currently do not have access to.


Our work provides valuable experimental insights into the spectral properties of the polariton laser emission and the condensate itself. We clearly observe energy fluctuations $\Delta E$ on the order of 1~GHz on timescales greater than 10$^{-3}$~s, independent of the experimental conditions. This drift can be minimised or controlled via active feedback using the AOM, similar to atomic condensates~\cite{thomsen_atom-laser_2002}. Faster feedback towards the GHz regime using an electro-optic modulator may result in an even smaller linewidth. \red{Electrical injection~\cite{bhattacharya2014room, schneider2013electrically} of polariton lasers can provide a suppression of the large $\Delta E$ at low frequencies since the pump and the cavity are embedded in one chip. However, it may also introduce a new set of noise. Further research in this area would be valuable. }

Furthermore, our work sets the ground for experimentally probing the role of interparticle interactions on the decoherence of polariton condensates~\cite{thomsen_atom-laser_2002,Whittaker_2009,amelio2022bogoliubov}. With our set of experimental parameters, we did not observe any clear transition from Lorentzian to Gaussian spectral distributions, predicted for stronger interactions~\cite{thomsen_atom-laser_2002}, suggesting that the interaction energy is too weak to observe this effect (due to the small excitonic fraction of $|X|^2=0.13$). We expect this transition to be revealed by performing similar experiments that combine excitation power series and increasing excitonic fraction, with additional measurements of the absolute value and noise strength of the number of polaritons~\cite{amelio2022bogoliubov}. This will be the subject of our future work.

\begin{backmatter}
\bmsection{Funding} Australian Research Council (CE170100039, DE220100712). National Science Centre (Poland)/Narodowe Centrum Nauki 2020/39/D/ST3/03546.

\bmsection{Acknowledgments} M.W acknowledges funding from Schmidt Science Fellows.

\bmsection{Disclosures} The authors declare no conflicts of interest.

\bmsection{Data Availability Statement} 
\red{Data underlying the results presented in this paper are not publicly available at this time but may be obtained from the authors upon reasonable request. }





\bmsection{Supplemental document}
See Supplement 1 for supporting content. 

\end{backmatter}

\bibliography{reference}

\begin{thebibliography}{10}
\newcommand{\enquote}[1]{``#1''}

\bibitem{schawlow_infrared_1958}
A.~L. Schawlow and C.~H. Townes, \enquote{Infrared and optical masers,} {\protect\JournalTitle{Physical Review}} \textbf{112}, 1940--1949.

\bibitem{thomsen_atom-laser_2002}
L.~K. Thomsen and H.~M. Wiseman, \enquote{Atom-laser coherence and its control via feedback,} {\protect\JournalTitle{Physical Review A}} \textbf{65}, 063607 (2002).

\bibitem{kasprzak2006bose}
J.~Kasprzak \emph{et~al.}, \enquote{Bose--{E}instein condensation of exciton polaritons,} {\protect\JournalTitle{{Nature} (London)}} \textbf{443}, 409--414 (2006).

\bibitem{Balili1007}
R.~Balili, V.~Hartwell, D.~Snoke, \emph{et~al.}, \enquote{Bose-{E}instein condensation of microcavity polaritons in a trap,} {\protect\JournalTitle{Science}} \textbf{316}, 1007--1010 (2007).

\bibitem{imamog1996nonequilibrium}
A.~Imamoglu, R.~Ram, S.~Pau, and Y.~Yamamoto, \enquote{Nonequilibrium condensates and lasers without inversion: Exciton-polariton lasers,} {\protect\JournalTitle{Physical Review A}} \textbf{53}, 4250 (1996).

\bibitem{deng2003polariton}
H.~Deng, G.~Weihs, D.~Snoke, \emph{et~al.}, \enquote{Polariton lasing vs. photon lasing in a semiconductor microcavity,} {\protect\JournalTitle{Proceedings of the National Academy of Sciences}} \textbf{100}, 15318--15323 (2003).

\bibitem{Zhang2022}
L.~Zhang, J.~Hu, J.~Wu, \emph{et~al.}, \enquote{Recent developments on polariton lasers,} {\protect\JournalTitle{Progress in Quantum Electronics}} \textbf{83}, 100399 (2022).

\bibitem{kavokin_baumberg_malpuech_laussy}
A.~Kavokin, J.~Baumberg, G.~Malpuech, and F.~Laussy, \emph{Polartion Devices} (Oxford University Press, 2017), pp. 519--546.

\bibitem{Sanvitto2016}
D.~Sanvitto and S.~K{\'e}na-Cohen, \enquote{The road towards polaritonic devices,} {\protect\JournalTitle{Nature Materials}} \textbf{15}, 1061 EP -- (2016).

\bibitem{kavokin2022polariton}
A.~Kavokin, T.~C. Liew, C.~Schneider, \emph{et~al.}, \enquote{Polariton condensates for classical and quantum computing,} {\protect\JournalTitle{Nature Reviews Physics}} \textbf{4}, 435--451 (2022).

\bibitem{tassone2000lasing}
F.~Tassone and Y.~Yamamoto, \enquote{Lasing and squeezing of composite bosons in a semiconductor microcavity,} {\protect\JournalTitle{Physical Review A}} \textbf{62}, 063809 (2000).

\bibitem{porras2003linewidth}
D.~Porras and C.~Tejedor, \enquote{Linewidth of a polariton laser: Theoretical analysis of self-interaction effects,} {\protect\JournalTitle{Physical Review B}} \textbf{67}, 161310 (2003).

\bibitem{amelio2022bogoliubov}
I.~Amelio and I.~Carusotto, \enquote{Bogoliubov theory of the laser linewidth and application to polariton condensates,} {\protect\JournalTitle{Physical Review A}} \textbf{105}, 023527 (2022).

\bibitem{Whittaker_2009}
D.~M. Whittaker and P.~R. Eastham, \enquote{Coherence properties of the microcavity polariton condensate,} {\protect\JournalTitle{{EPL} (Europhysics Letters)}} \textbf{87}, 27002 (2009).

\bibitem{byrnes2014exciton}
T.~Byrnes, N.~Y. Kim, and Y.~Yamamoto, \enquote{Exciton--polariton condensates,} {\protect\JournalTitle{Nature Physics}} \textbf{10}, 803--813 (2014).

\bibitem{Love2008}
A.~P.~D. Love, D.~N. Krizhanovskii, D.~M. Whittaker, \emph{et~al.}, \enquote{Intrinsic decoherence mechanisms in the microcavity polariton condensate,} {\protect\JournalTitle{Phys. Rev. Lett.}} \textbf{101}, 067404 (2008).

\bibitem{Kim2016}
S.~Kim, B.~Zhang, Z.~Wang, \emph{et~al.}, \enquote{Coherent polariton laser,} {\protect\JournalTitle{Phys. Rev. X}} \textbf{6}, 011026 (2016).

\bibitem{Askitopoulos2019}
A.~Askitopoulos, L.~Pickup, S.~Alyatkin, \emph{et~al.}, \enquote{Giant increase of temporal coherence in optically trapped polariton condensate,} {\protect\JournalTitle{arXiv preprint arXiv:1911.08981}}  (2019).

\bibitem{Wouters2008}
M.~Wouters, I.~Carusotto, and C.~Ciuti, \enquote{Spatial and spectral shape of inhomogeneous nonequilibrium exciton-polariton condensates,} {\protect\JournalTitle{Phys. Rev. B}} \textbf{77}, 115340 (2008).

\bibitem{ostrovskaya2012dissipative}
E.~A. Ostrovskaya, J.~Abdullaev, A.~S. Desyatnikov, \emph{et~al.}, \enquote{Dissipative solitons and vortices in polariton bose-einstein condensates,} {\protect\JournalTitle{Physical Review A}} \textbf{86}, 013636 (2012).

\bibitem{roumpos2010gain}
G.~Roumpos, W.~H. Nitsche, S.~H{\"o}fling, \emph{et~al.}, \enquote{Gain-induced trapping of microcavity exciton polariton condensates,} {\protect\JournalTitle{Physical Review Letters}} \textbf{104}, 126403 (2010).

\bibitem{orfanakis_ultralong_2021}
K.~Orfanakis, A.~F. Tzortzakakis, D.~Petrosyan, \emph{et~al.}, \enquote{Ultralong temporal coherence in optically trapped exciton-polariton condensates,} {\protect\JournalTitle{Physical Review B}} \textbf{103}, 235313 (2021).

\bibitem{sigurdsson_persistent_2022}
H.~Sigurdsson, I.~Gnusov, S.~Alyatkin, \emph{et~al.}, \enquote{Persistent {Self}-{Induced} {Larmor} {Precession} {Evidenced} through {Periodic} {Revivals} of {Coherence},} {\protect\JournalTitle{Physical Review Letters}} \textbf{129}, 155301 (2022).

\bibitem{barrat_qubit_2023}
J.~Barrat, A.~F. Tzortzakakis, M.~Niu, \emph{et~al.}, \enquote{Qubit {Analog} with {Polariton} {Superfluid} in an {Annular} {Trap},}  (2023). ArXiv:2308.05555 [cond-mat, physics:physics, physics:quant-ph].

\bibitem{kuznetsov2023microcavity}
A.~S. Kuznetsov, K.~Biermann, A.~A. Reynoso, \emph{et~al.}, \enquote{Microcavity phonoritons--a coherent optical-to-microwave interface,} {\protect\JournalTitle{Nature Communications}} \textbf{14}, 5470 (2023).

\bibitem{askitopoulos2013polariton}
A.~Askitopoulos, H.~Ohadi, A.~V. Kavokin, \emph{et~al.}, \enquote{Polariton condensation in an optically induced two-dimensional potential,} {\protect\JournalTitle{Phys. Rev. B}} \textbf{88}, 041308 (2013).

\bibitem{estrecho2019direct}
E.~Estrecho, T.~Gao, N.~Bobrovska, \emph{et~al.}, \enquote{Direct measurement of polariton-polariton interaction strength in the {T}homas-{F}ermi regime of exciton-polariton condensation,} {\protect\JournalTitle{Phys. Rev. B}} \textbf{100}, 035306 (2019).

\bibitem{sun2017direct}
Y.~Sun, Y.~Yoon, M.~Steger, \emph{et~al.}, \enquote{Direct measurement of polariton--polariton interaction strength,} {\protect\JournalTitle{Nature Physics}} \textbf{13}, 870--875 (2017).

\bibitem{SunPolLifetime}
Y.~Sun, P.~Wen, Y.~Yoon, \emph{et~al.}, \enquote{Bose-{E}instein condensation of long-lifetime polaritons in thermal equilibrium,} {\protect\JournalTitle{Phys. Rev. Lett.}} \textbf{118}, 016602 (2017).

\bibitem{alnatah2024coherence}
H.~Alnatah, Q.~Yao, J.~Beaumariage, \emph{et~al.}, \enquote{Coherence measurements of polaritons in thermal equilibrium reveal a power law for two-dimensional condensates,} {\protect\JournalTitle{Science Advances}} \textbf{10}, eadk6960 (2024).

\bibitem{pieczarka2022bogoliubov}
M.~Pieczarka, O.~Bleu, E.~Estrecho, \emph{et~al.}, \enquote{Bogoliubov excitations of a polariton condensate in dynamical equilibrium with an incoherent reservoir,} {\protect\JournalTitle{Physical Review B}} \textbf{105}, 224515 (2022).

\bibitem{bieganska2021collective}
D.~Biega{\'n}ska, M.~Pieczarka, E.~Estrecho, \emph{et~al.}, \enquote{Collective excitations of exciton-polariton condensates in a synthetic gauge field,} {\protect\JournalTitle{Physical Review Letters}} \textbf{127}, 185301 (2021).

\bibitem{pieczarka2020observation}
M.~Pieczarka, E.~Estrecho, M.~Boozarjmehr, \emph{et~al.}, \enquote{Observation of quantum depletion in a non-equilibrium exciton--polariton condensate,} {\protect\JournalTitle{Nature Communications}} \textbf{11}, 429 (2020).

\bibitem{estrecho2021low}
E.~Estrecho, M.~Pieczarka, M.~Wurdack, \emph{et~al.}, \enquote{Low-energy collective oscillations and bogoliubov sound in an exciton-polariton condensate,} {\protect\JournalTitle{Physical Review Letters}} \textbf{126}, 075301 (2021).

\bibitem{alnatah2024critical}
H.~Alnatah, P.~Comaron, S.~Mukherjee, \emph{et~al.}, \enquote{Critical fluctuations in a confined driven-dissipative quantum condensate,} {\protect\JournalTitle{Science Advances}} \textbf{10}, eadi6762 (2024).

\bibitem{Steger_polaritonlifetime}
M.~Steger, G.~Liu, B.~Nelsen, \emph{et~al.}, \enquote{Long-range ballistic motion and coherent flow of long-lifetime polaritons,} {\protect\JournalTitle{Phys. Rev. B}} \textbf{88}, 235314 (2013).

\bibitem{KRIZHANOVSKII2001583}
D.~Krizhanovskii, A.~Tartakovskii, A.~Chernenko, \emph{et~al.}, \enquote{Energy relaxation of resonantly excited polaritons in semiconductor microcavities,} {\protect\JournalTitle{Solid State Communications}} \textbf{118}, 583--587 (2001).

\bibitem{Tassone1997}
F.~Tassone, C.~Piermarocchi, V.~Savona, \emph{et~al.}, \enquote{Bottleneck effects in the relaxation and photoluminescence of microcavity polaritons,} {\protect\JournalTitle{Phys. Rev. B}} \textbf{56}, 7554--7563 (1997).

\bibitem{loudon2000quantum}
R.~Loudon, \emph{The quantum theory of light} (OUP Oxford, 2000).

\bibitem{Pieczarka2019}
M.~Pieczarka, M.~Boozarjmehr, E.~Estrecho, \emph{et~al.}, \enquote{Effect of optically induced potential on the energy of trapped exciton polaritons below the condensation threshold,} {\protect\JournalTitle{Phys. Rev. B}} \textbf{100}, 085301 (2019).

\bibitem{deng2010exciton}
H.~Deng, H.~Haug, and Y.~Yamamoto, \enquote{Exciton-polariton bose-einstein condensation,} {\protect\JournalTitle{Rev. Mod. Phys.}} \textbf{82}, 1489 (2010).

\bibitem{bai2021narrow}
Z.~Bai, Z.~Zhao, Y.~Qi, \emph{et~al.}, \enquote{Narrow-linewidth laser linewidth measurement technology,} {\protect\JournalTitle{Frontiers in Physics}} p. 684 (2021).

\bibitem{estrecho2018single}
E.~Estrecho, T.~Gao, N.~Bobrovska, \emph{et~al.}, \enquote{Single-shot condensation of exciton polaritons and the hole burning effect,} {\protect\JournalTitle{Nature Communications}} \textbf{9}, 2944 (2018).

\bibitem{serkland2020narrow}
D.~K. Serkland, T.~J. Morin, H.~M. So, \emph{et~al.}, \enquote{Narrow-linewidth vcsels based on multi-mirror cavities (conference presentation),} in \emph{Vertical-Cavity Surface-Emitting Lasers XXIV,}  vol. 11300 (SPIE, 2020), p. 1130008.

\bibitem{huang2022narrow}
M.~Huang, D.~Serkland, and J.~Camparo, \enquote{A narrow-linewidth three-mirror vcsel for atomic devices,} {\protect\JournalTitle{Applied Physics Letters}} \textbf{121} (2022).

\bibitem{bhattacharya2014room}
P.~Bhattacharya, T.~Frost, S.~Deshpande, \emph{et~al.}, \enquote{Room temperature electrically injected polariton laser,} {\protect\JournalTitle{Physical review letters}} \textbf{112}, 236802 (2014).

\bibitem{schneider2013electrically}
C.~Schneider, A.~Rahimi-Iman, N.~Y. Kim, \emph{et~al.}, \enquote{An electrically pumped polariton laser,} {\protect\JournalTitle{Nature}} \textbf{497}, 348--352 (2013).

\bibitem{xue2021split}
Y.~Xue, I.~Chestnov, E.~Sedov, \emph{et~al.}, \enquote{Split-ring polariton condensates as macroscopic two-level quantum systems,} {\protect\JournalTitle{Physical Review Research}} \textbf{3}, 013099 (2021).

\bibitem{Haug}
H.~Haug, H.~T. Cao, and D.~B.~T. Thoai, \enquote{Coherence and decoherence of a polariton condensate,} {\protect\JournalTitle{Phys. Rev. B}} \textbf{81}, 245309 (2010).

\end{thebibliography}


\end{document}